\documentclass[10pt,preprint,showpacs,nofootinbib,prd]{revtex4}

\usepackage{graphicx}% Include figure files
\usepackage{amsmath}
\usepackage{amssymb}
\usepackage{dcolumn}% Align table columns on decimal point
\usepackage{bm}% bold math
\usepackage{hyperref}

\begin{document}

%\preprint{APS/123-QED}

\title{Nonleptonic charmless two-body $B \to AT$ decays}% Force line breaks with \\

\author{J. H. Mu\~noz}
\email{jhmunoz@ut.edu.co}
\author{N. Quintero}
\email{nquinte@gmail.com}
\affiliation{Departamento de F\'isica, Universidad del Tolima A. A. 546, Ibagu\'e,
Colombia \vspace{5cm}}

%\date{}% It is always \today, today,
\vspace{7cm}
\begin{abstract}
\vspace{0.3cm}

In this work we have studied hadronic charmless
 two-body $B$ decays involving $p$-wave mesons in final state. We  have calculated
 branching ratios of  $B\to AT$ decays
(where $A$ and $T$ denotes a  $^3P_1$ axial-vector and a tensor meson, respectively),
using $B \to T$ form factors obtained in the covariant light-front (CLF) approach, and the full effective Hamiltonian. We have obtained that $\mathcal{B}(B^{0} \to a_{1}^{+}a_{2}^{-}) =42.47 \times10^{-6}$, $\mathcal{B}(B^{+} \to a_{1}^{+}a_{2}^{0}) = 22.71 \times10^{-6}$, $\mathcal{B}(B \to f_{1}K_{2}^{*}) = (2.8-4) \times 10^{-6}$  (with $f_{1}=, f_{1}(1285),f_{1}(1420)$) for $\theta_{^{3}P_{1}} = 53.2^{\circ}$, $\mathcal{B}(B \to f_{1}(1420)K_{2}^{*}) = (5.91-6.42) \times 10^{-6}$ with $\theta_{^{3}P_{1}} = 27.9^{\circ}$, $\mathcal{B}(B \to K_{1}a_{2})= (1.7 - 5.7) [1-9.3]  \times10^{-6}$ for $\theta_{K_{1}} =  -37^{\circ} [-58^{\circ}]$ where $K_1 = K_1(1270), K_1(1400)$. It seems that these decays  can be measured in experiments at $B$ factories. Additionally, we have found that   $\mathcal{B}(B \to K_{1}(1270)a_{2})/\mathcal{B}(B \to K_{1}(1400)a_{2})$ and $\mathcal{B}(B \to f_1(1420)K_{2}^{*})/\mathcal{B}(B \to f_1(1285)K_{2}^{*})$ ratios could be useful to determine  numerical values of  mixing angles $\theta_{K_{1}}$ and $\theta_{^{3}P_{1}}$, respectively.
\end{abstract}


\maketitle

%*********************************************************************************************
%*********************************************************************************************
%*********************************************************************************************

\section{Introduction}

Weak nonleptonic two-body  $B$ decays is a good  scenario to understand the interplay of short- and long-distance QCD effects,  to  investigate about CP violation, to test  some  QCD-motivated theories such as QCD factorization, perturbative QCD and soft-collinear    effective theory, to study physics beyond the Standard Model (SM), and constraint  numerical values of the Cabibbo-Kobayashi-Maskawa (CKM) parameters. \\

Hadronic charmless two-body $B \to M_1M_2$ decays, where $M_{1,2}$ can be a $l=0$ or a $l=1$ meson, have been broadly considered in the literature (for a recent review see Ref. \cite{Cheng09}). There are comprehensive and systematic articles about exclusive charmless $B \to PP, PV, VV$ (see for example Refs. \cite{Ali98,Chen99,adittional}), $ B \to PA, VA, AA$ \cite{axial,Calderon,Yang07,axialQCDF1,axialQCDF2}, $B \to SP, SV$  \cite{scalar},  $B \to PT, VT$ \cite{tensor1,tensor2}  ($P$, $V$, $S$, $A$ and $T$ denotes a pseudoscalar, a vector, a scalar, an axial-vector, and a tensor meson, respectively) decays. In this work,  we  have studied nonleptonic charmless two-body $B$ decays considering that both mesons in final state are orbitally excited $l=1$ mesons (or $p$-wave mesons). Specifically, we have  worked with $B \to AT$ decays, which could compete with $B \to VT$ modes and their branching fractions could be measured in the future LHCb experiment in the Large Hadron Collider (LHC) and at $B$-factories. Moreover, $B \to AT$ decays  can also offer a good place to study polarization in  a similar way to the   $B  \to VT$  scenario \cite{polarization}.\\

At experimental level, there are some recent measurements about  the production of tensor or axial-vector mesons in $B$ decays. Recently,  BABAR Collaboration reported  branching fractions of nonleptonic charmless two-body $B$ decays involving tensor mesons   in final state \cite{BaBartensor}. On the other hand, hadronic charmless  $B^{0} \to a_{1}(1260)^{\pm}\pi^{\mp}$ decays were the first modes with axial-vector mesons in final state, measured by both $B$ factories, BABAR \cite{BABAR1} and Belle \cite{Belle}. Additionally, BABAR Collaboration  reported the observation of other decays with  axial-vector mesons in final state \cite{BaBaraxial}. In general, these modes have branchings of the order of $10^{-6}$. So far, there is not experimental information about $B \to AT$ decays.  \\

In this work we extend the knowledge of weak nonleptonic two-body $B$ decays, considering that  both mesons in final state are $p$-wave. In principle, we have six possibilities for considering two orbitally excited (or $p$ wave) mesons in final state: $B \to S(S, A, T)$, $B \to A(A, T)$ and $B \to TT$. In this paper, we have computed branching ratios of exclusive charmless $B \to AT$ (where $A$ is a $^3P_1$ axial-vector meson) decays assuming generalized factorization, considering the  effective weak Hamiltonian $H_{eff}$ and taking  $B \to T$ form factors from the covariant-light front (CLF)  approach \cite{Cheng04}, which is one of the few models that provides the evaluation of the hadronic matrix element $\langle T|J_{\mu}|B \rangle$.\\

This paper is organized as follows: in Sec. II, we discuss briefly about the effective weak Hamiltonian, factorization scheme and $B \rightarrow T$ form factors in the CLF approach. Sec. III is dedicated to describe input parameters. In Sec. IV, we present our numerical results for  branching fractions and  conclusions are given in Sec. V. In Appendix, we display explicitly  decay amplitudes for charmless $B \to A(^3P_1)T$ modes. \\

%*********************************************************************************************
%*********************************************************************************************
%*********************************************************************************************

\section{Theoretical framework}

\subsection{The weak effective   Hamiltonian and factorization approach}

The weak  effective  Hamiltonian $H_{eff}$ for nonleptonic charmless  two-body $B$ decays is \cite{buras96}:

\begin{eqnarray}
H_{eff}(\Delta B=1) &=& \frac{G_F}{\sqrt{2}}\Bigg[V_{ub}V^*_{uq} \Big(C_1(\mu)
O^u_1(\mu) + C_2(\mu) O^u_2(\mu)\Big) + V_{cb}V^*_{cq}\Big(C_1(\mu)
O^c_1(\mu) + C_2(\mu) O^c_2(\mu)\Big)  \nonumber \\
&&- V_{tb}V^*_{tq}\Bigg(\sum^{10}_{i=3}C_i(\mu) O_i(\mu) \Bigg) \Bigg] + h.c.\ ,
\end{eqnarray}

\noindent where $G_F$ denotes the Fermi constant, $C_i(\mu)$ are Wilson
coefficients evaluated at renormalization scale $\mu$ and  coefficients $V_{mn}$  are CKM matrix elements related to the transition. Local
operators $O_i$ are given by

\begin{itemize}
  \item current-current (tree) operators
  \begin{eqnarray}
  O^u_1 &=& (\bar q_\alpha u_\alpha)_{V-A} \cdot (\bar u_\beta b_\beta)_{V-A}  \nonumber \\
O^u_2 &=& (\bar q_\alpha u_\beta)_{V-A} \cdot (\bar u_\beta b_\alpha)_{V-A}  \nonumber \\
  O^c_1 &=& (\bar q_\alpha c_\alpha)_{V-A} \cdot (\bar c_\beta b_\beta)_{V-A}  \nonumber \\
O^c_2 &=& (\bar q_\alpha c_\beta)_{V-A} \cdot (\bar c_\beta b_\alpha)_{V-A}  \nonumber \\
\end{eqnarray}
  \item QCD penguin operators
  \begin{eqnarray}
  O_{3(5)} &=& (\bar q_\alpha b_\alpha)_{V-A} \cdot \sum_{q^{\prime }}
(\bar q^{\prime }_\beta q^{\prime }_\beta)_{V-A(V+A)}  \nonumber \\
O_{4(6)} &=& (\bar q_\alpha b_\beta)_{V-A} \cdot \sum_{q^{\prime }}
(\bar q^{\prime }_\beta q^{\prime }_\alpha)_{V-A(V+A)}  \nonumber \\
\end{eqnarray}
  \item electroweak penguin operators
  \begin{eqnarray}
  O_{7(9)} &=& \frac{3}{2}(\bar q_\alpha b_\alpha)_{V-A} \cdot
\sum_{q^{\prime }} e_{q^{\prime }}(\bar q^{\prime }_\beta q^{\prime }_\beta)_{V+A(V-A)} \nonumber \\
O_{8(10)} &=& \frac{3}{2}(\bar q_\alpha b_\beta)_{V-A} \cdot
\sum_{q^{\prime }} e_{q^{\prime }} (\bar q^{\prime }_\beta q^{\prime }_\alpha)_{V+A(V-A)},
\end{eqnarray}
\end{itemize}

\noindent where $(\bar q_{1}q_{2})_{V\mp A} \equiv \bar q_{1}\gamma_\mu(1\mp\gamma_5)q_{2}$, and symbols $\alpha$ and $\beta$ are $SU(3)$ color indices. The sums run over  active quarks at the scale $\mu=\mathcal{O}(m_b)$, i.e. $%
q^{\prime }$ runs with quarks $u$, $d$, $s$ and $c$.\\

In order to obtain  branching ratios of nonleptonic two-body  $B \rightarrow M_1M_2$ decays it is necessary to evaluate the hadronic matrix element involving four-quark operators $\langle M_{1}M_{2}|O_i|B \rangle$. In the framework of factorization approach, it can be approximated  by the product of two matrix elements of single currents: $\langle M_{1}|(J_1)_{\mu}|0\rangle \langle M_{2}|(J_2)^{\mu}|B \rangle$ or $\langle M_{2}|(J_1)_{\mu}|0\rangle \langle M_{1}|(J_2)^{\mu}|B \rangle$, where $J_{\mu}$ is a bilinear current. Thus, the matrix element of a four-quark operator is expressed  as the product of decay constant and form factors. The hadronic matrix element is renormalization scheme and scale independent \cite{Buras} while Wilson Coefficients are renormalization scheme and scale dependent. So, the decay amplitude under this naive factorization is not truly  physical. \\

The improved generalized factorization  solves the aforementioned scale problem.  For example, in Refs. \cite{Ali98, Chen99} it is considered a method to extract the $\mu$ dependence from the matrix element $\langle O_i(\mu) \rangle$ and combine it with the $\mu$-dependent Wilson coefficients $C_i(\mu)$ to form $\mu$-independent effective Wilson coefficients $c_i^{eff}$. We have taken  the respective numerical values for $c_i^{eff}$ reported in Table I of Ref. \cite{Calderon}. They were calculated in next to leading order  Wilson coefficients for $\Delta B=1$
transitions obtained in the naive dimensional regularization  scheme. \\

Effective Wilson coefficients $c_i^{eff}$ appear in factorizable decay amplitudes  as linear combinations. It allows to define effective coefficients $a_i$, which are renormalization scale and scheme independent. $a_i$'s are defined as:

\begin{eqnarray}
a_i &\equiv& c^{eff}_i + \frac{1}{N_c} c^{eff}_{i+1}\ (i=odd)  \nonumber \\
a_i &\equiv& c^{eff}_i + \frac{1}{N_c} c^{eff}_{i-1}\ (i=even)\ ,
\end{eqnarray}

\noindent where the index $i$ runs over ($1,...,10$) and $N_c=3$ is the
 number of colors. Phenomenologically,   nonfactorizable contributions to the hadronic matrix element are modeled by treating $N_c$ as a free parameter and its value can be extracted from experiment. In this work we have used  numerical values for  $a_i$ coefficients for $b \rightarrow d$ and $b \rightarrow s$ transitions displayed in Table II of Ref. \cite{Calderon}. \\

%*********************************************************************************************
%*********************************************************************************************
%*********************************************************************************************

\subsection{Form Factors in the CLF approach}

In the framework of generalized factorization the hadronic matrix element $\langle AT|O_i|B \rangle$ is approximated by $\langle A|(J_1)_{\mu}|0\rangle \langle T|(J_2)^{\mu}|B \rangle$.  So, we need to compute  the hadronic matrix element $\langle T|J^{\mu}|B \rangle$ in order to obtain numerical values for branching ratios of $B \rightarrow AT$ decays. We have used the parametrization given in Ref. \cite{ISGW}:

\begin{eqnarray}
  \langle T|V^{\mu}|B\rangle &=& i h(q^{2}) \varepsilon^{\mu\nu\rho\sigma} \epsilon_{\nu\alpha}p_{B}^{\alpha} (p_{B}+p_{T})_{\rho}(p_{B}-p_{T})_{\sigma}, \nonumber\\
  \langle T|A^{\mu}|B\rangle &=& k(q^{2}) \epsilon^{*\mu\nu} (p_{B})_{\nu}+ \epsilon_{\alpha\beta}^{*}p_{B}^{\alpha}p_{B}^{\beta} \nonumber\\
  & & \left[ b_{+}(q^{2})(p_{B}+p_{T})^{\mu} + b_{-}(q^{2})(p_{B}-p_{T})^{\mu} \right],
\end{eqnarray}

\noindent where $V^{\mu}$ and $A^{\mu}$ denote the vector and the axial-vector current, respectively;  $\epsilon_{\mu\nu}$ is the polarization  of tensor meson, $p_{B}$ and $p_{T}$ are the momentum of $B$ and $T$ mesons, respectively, and  $h, k, b_{\pm}$ are  form factors for the $B \rightarrow T$ transition.\\

So far,  only two models\footnote{Recently, Ref. \cite{LEET} calculated $B \to K_2^*$ form factors  using large energy effective theory (LEET) techniques.}  provide a systematical estimate of $B \rightarrow T$ form factors: the ISGW model \cite{ISGW} with its improved version \cite{ISGW2} and  the CLF quark model \cite{Jaus}.  Keeping in mind that the improved ISGW2 model \cite{ISGW2}  has difficulties in the low-$q^2$ region, in particular, at the maximum $q^2=0$ recoil point where the final-state meson could be highly relativistic, we have  used numerical values for form factors $h, k, b_{\pm}$,  obtained in the CLF quark model \cite{Cheng04}. This reference extended the covariant analysis of the light-front approach \cite{Jaus} to even-parity, $p$-wave mesons. \\

 A light-front quark model (LFQM) provides a relativistic study of the movement of the hadron and also gives a fully description of the hadron spin. The light-front wave functions do not depend on   the hadron momentum and  are explicitly Lorentz invariant. In the CLF quark model, the spurious contribution, which  is dependent of   the orientation of the light-front, is cancelled by  inclusion of the zero mode contribution, and becomes irrelevant in decay constants and form factors, so that the result is guaranteed to be covariant and more self consistent. \\

This model has been used by different authors in the last five years, obtaining, in some cases, predictions that are favorable with available experimental information.  For example, some authors worked with semileptonic decays of $B_c$ meson including $s$- and  $p$-wave mesons in  final state  \cite{Lu(Bc)} and nonleptonic $B_c^- \rightarrow X(3872)\pi^-(K^-)$  modes \cite{Lu(Bc)07}. Others, studied two-photon  annihilation $P \rightarrow \gamma \gamma$ and magnetic dipole transition $V \rightarrow P\gamma$ processes for the ground-state heavy quarkonium within the CLF approach \cite{quarkonium}, and   radiative $B \rightarrow (K^*, K_1, K_2^*)\gamma$ channels in the same framework \cite{Radiative}. Ref. \cite{Chen07} examined $B \rightarrow (K_{0}^{*}(1430), K_{2}^{*}(1430))\phi$ in the LFQM. Recently, we computed branching ratios of hadronic charmless $B \to P(V)T$ decays in the CLF approach \cite{tensor1}. \\

In the CLF approach, form factors are explicit functions of $q^2$ in the space-like region and then analytically extend them to the time-like region in order to compute physical form factors at  $q^{2}\geq 0$.   They are parametrized and reproduced in the three-parameter form \cite{Cheng04}:

\begin{equation}\label{1}
F(q^{2}) = \frac{F(0)}{1-aX+bX^{2}},
\end{equation}

\noindent with $X=q^{2}/m_{B}^{2}$.  In Tables VI and VII of Ref. \cite{Cheng04} it is displayed the parameters $a$, $b$ and $F(0)$ (form factor at the zero momentum transfer) for   $B \rightarrow a_2(1320)$ and  $B \rightarrow K_2^*(1430)$ transitions, which are $B \rightarrow T$ transitions required in this work. In Table I,  we have summarized these numerical values.\\

\begin{table}[ht]
{\small Table I.~ Form factors for $B \rightarrow a_{2}(1320)$ and $B \rightarrow K_{2}^{*}(1430)$ transitions obtained in the CLF model \cite{Cheng04}  are fitted to the 3-parameter form in Eq.(\ref{1}).  $k$ is  dimensionless and  $h$, $b_+$, $b_-$ have dimensions of GeV$^{-2}$.}
\par
\begin{center}
\renewcommand{\arraystretch}{1.2}
\renewcommand{\arrayrulewidth}{0.8pt}
\begin{tabular}{l|ccc|ccc}
\hline\hline
 & & $B \rightarrow a_{2}$ & & & $B \rightarrow K_{2}^{*}$ &\\
 \hline
$F$ & $F(0)$ & $a$ & $b$ &  $F(0)$ & $a$ & $b$\\
\hline
$h$ &  0.008 & 2.20 & 2.30 &   0.008 & 2.17 & 2.22\\
   $k$ &  0.031 & $-2.47$ & $2.47$ &   0.015 & $-3.70$ & 1.78\\
   $b_{+}$ &  $-0.005$ & 1.95 & 1.80 &   $-0.006$ & 1.96 & 1.79\\
   $b_{-}$ &  0.0016 & $-0.23$ & 1.18 &  0.002 & 0.38 & 0.92\\
\hline\hline
\end{tabular}%
\end{center}
\end{table}

The matrix element of the current between the vacuum and final  $^3P_1$ axial-vector  meson ($A$) can be expressed in terms of the respective decay constants $f_{A}$, in the form

\begin{eqnarray}
\langle A(p_A,\epsilon)|A_\mu| 0\rangle &=& f_{A}m_{A} \epsilon_\mu,
\end{eqnarray}

\noindent where $\epsilon_\mu$ is the vector polarization of the $^3P_1$ axial-vector  meson. On the other hand, the polarization $\epsilon_{\mu\nu}$ of the $^{3}P_{2}$ tensor meson satisfies the relations

\begin{equation}\label{}
\epsilon_{\mu\nu}=\epsilon_{\nu\mu}, \qquad \epsilon_{\mu}^{\mu}=0, \qquad p_{\mu}\epsilon^{\mu\nu}=p_{\nu}\epsilon^{\mu\nu}=0.
\end{equation}

\noindent Therefore,

\begin{equation}\label{}
\langle 0|(V-A)_{\mu}|T\rangle = a\epsilon_{\mu\nu}p^{\nu}+ b \epsilon^{\nu}_{\nu}p_{\mu}=0,
\end{equation}

\noindent and hence the decay constant of the tensor meson vanishes, i.e., the tensor meson can not be produced from the vacuum and we can not approximate the hadronic matrix element $\langle AT|O_i|B \rangle$  by $\langle T|(J_1)_{\mu}|0\rangle \langle A|(J_2)^{\mu}|B \rangle$. This fact simplifies considerably decay amplitudes for $B \rightarrow A(^3P_1)T$ processes if we compare them with those for charmless two-body $B$ decays such as $B \to PP,\; PV$, and $VV$ \cite{Ali98, Chen99} and $B \to AP,\; AV$, and $AA$ \cite{axial,Calderon}.\\

On the other hand, we do not have considered $B \to MT$ decays where $M$ can be a $^1P_1$ axial-vector or a  scalar meson. If we assume  factorization hypothesis, the amplitude decay of these modes is $\langle M|(J_1)_{\mu}|0\rangle \langle T|(J_2)^{\mu}|B \rangle$. There is not a contribution of the form $\langle T|(J_1)_{\mu}|0\rangle \langle M|(J_2)^{\mu}|B \rangle$ because  $\langle T |J_{\mu}|0 \rangle$ is zero. Thus, $B \to MT$ decays, in general, imply the evaluation of  $\langle M|J_{\mu}|0\rangle$.\\

$^1P_1$ axial-vector mesons  with $J^{PC}=J^{+-}$ ($b_1$ and  $h_1$) have even $G$-parity and the  axial current which produces a $b_1$ or a $h_1$ meson has odd $G$-parity. So, $\langle 0| \bar{u} \gamma_\mu \gamma_5 d | b_1 (h_1)\rangle = 0$ by $G$-parity conservation and hence $f_{b_1} = f_{h_1} = 0$. In other words, for the $^1P_1$ axial-vector meson  its decay constant is small, vanishing in the $SU(3)$ limit.  So, we do not consider in this work $B \to A(^1P_1)T$ decays because their factorizable amplitude  is proportional to decay constant $f_{A(^1P_1)}$. \\

The situation with hadronic $B \to ST$ decays is similar. The vector decay constant of scalar mesons,  defined as $\langle S(p)| \bar{q_i} \gamma_\mu q_j | 0 \rangle = f_Sp_{\mu}$, is either zero or small (of order of  $m_d - m_u$, $m_s - m_{d,u}$). Moreover, the identification of light scalar mesons is not easy experimentally and the underlying structure is not well understood at theoretical level \cite{Cheng09}. For these reasons, we have not studied neither $B \to ST$ decays in this work.\\

%*********************************************************************************************
%*********************************************************************************************
%*********************************************************************************************

\section{Input parameters}

In this section we present  numerical inputs that are necessary to obtain our predictions.  We also discuss about mixing angles between $K_{1A}$ and $K_{1B}$ mesons  and    ${}^{3}P_{1}$ states $f_{1}(1285)$ and $f_{1}(1420)$.\\

In the quark model, there are two nonets of $J^{P}=1^{+}$ axial-vector
mesons  as the orbital excitation of the $q\bar{q}$
system. In terms of the spectroscopic notation $^{2S+1}L_{J}$,
these two types of axial-vector mesons are  $^{3}P_{1}$
and $^{1}P_{1}$. They have distinctive $C$ quantum
numbers, $C=+$ and $C=-$, respectively. Experimentally,
the $J^{PC}=1^{++}$ nonet consists of $a_{1}(1260)$,
$f_{1}(1285)$, $f_{1}(1420)$, and $K_{1A}$, while the $1^{+-}$ nonet is conformed by
$b_{1}(1235)$, $h_{1}(1170)$, $h_{1}(1380)$, and $K_{1B}$ \cite{PDG}. However, the physical strange axial-vector mesons $K_{1}(1270)$ and $K_{1}(1400)$ are a mixture of $K_{1A}$ and $K_{1B}$:
\begin{equation}\label{}
\begin{array}{c}
  K_{1}(1270) = K_{1A}\sin\theta_{K_{1}} + K_{1B}\cos\theta_{K_{1}}\\
  K_{1}(1400) = K_{1A}\cos\theta_{K_{1}} - K_{1B}\sin\theta_{K_{1}},
\end{array}
\end{equation}

\noindent where $\theta_{K_{1}}$ is the $K_{1A} - K_{1B}$ mixing angle.\\

We used two different set of mixing angle predictions given in  Ref. \cite{axialQCDF2}: $\theta_{K_{1}} = - 37^{\circ}, - 58^{\circ}$, where $\theta_{K_{1}}$ is favored to be negative as implied by the experimental measurement of the  $\mathcal{B}(B \to K_{1}(1270)\gamma)/\mathcal{B}(B \to K_{1}(1400)\gamma)$ ratio in $B$ decays.  In Table II, we present numerical values of decay constants depending of the value of $\theta_{K_{1}}$. Additionally,  Ref. \cite{Hatanaka} predicted  that the mixing angle $\theta_{K_{1}}$ must be negative, $\theta_{K_{1}} = -34^{\circ}$ and obtained $f_{K_{1}(1270)}$ and $f_{K_{1}(1400)}$ (see Table II), from the combining analysis for  $B \to K_{1}\gamma$ and $\tau^{-} \to K_{1}(1270)^{-}\nu_{\tau}$ decays. In this work,  the $K_{1A}-K_{1B}$ mixing is introduced through decay constants.\\

\begin{table}[ht]
{\small Table II. Numerical values (in MeV) of decay constant $f_{K_{1}}$.}
\par
\begin{center}
\renewcommand{\arraystretch}{1.2}
\renewcommand{\arrayrulewidth}{0.8pt}
\begin{tabular}{lcc}
\hline\hline
 $\theta_{K_{1}}$ & $f_{K_{1}(1270)}$  & $f_{K_{1}(1400)}$  \\
 \hline
  $- 37^{\circ}$ \cite{axialQCDF2}  & ($-184$ $\pm$ 11) &  (177 $\pm$ 12) \\
  $- 58^{\circ}$ \cite{axialQCDF2} & ($-234$ $\pm$ 11) & (100 $\pm$ 12) \\
  $- 34^{\circ}$ \cite{Hatanaka} & ($-169$ $\pm$ 25) & (179 $\pm$ 13) \\
\hline\hline
\end{tabular}%
\end{center}
\end{table}

Analogous to the $\eta-\eta'$ mixing in the pseudoscalar
nonet,  ${}^{3}P_{1}$ states $f_{1}(1285)$ and $f_{1}(1420)$ are mixed in terms of the pure octet $|f_8\rangle$ and singlet $|f_1\rangle$ due to $SU(3)$ breaking effects, and can be parameterized as \cite{axialQCDF1,mixing1}:
\begin{eqnarray}
&& |f_{1}(1285) \rangle = |f_1\rangle \cos \theta_{^{3}P_{1}}  + |f_8\rangle \sin \theta_{^{3}P_{1}},
  \nonumber \\
&& |f_{1}(1420)\rangle = - |f_1\rangle \sin \theta_{^{3}P_{1}} + |f_8\rangle \cos \theta_{^{3}P_{1}}.
\end{eqnarray}

\noindent Decay constants $f^{q}_{f_{1}(1285)}$ and $f^{q}_{f_{1}(1420)}$ are defined  by

\begin{eqnarray}
&& \langle 0|\bar q \gamma_\mu \gamma_5 q| f_{1}(1285)\rangle = -i m_{f_{1}(1285)} f^{q}_{f_{1}(1285)} \epsilon_{\mu},
  \nonumber \\
&& \langle 0|\bar q \gamma_\mu \gamma_5 q| f_{1}(1420)\rangle = -i m_{f_{1}(1420)} f^{q}_{f_{1}(1420)} \epsilon_{\mu}.
\end{eqnarray}

\noindent Thus, it is obtained
\begin{eqnarray}
&& f^{u}_{f_{1}(1285)} = \frac{f_{f_1}}{\sqrt{3}} \frac{m_{f_1}}{ m_{f_{1}(1285)}}\cos\theta_{^{3}P_{1}} + \frac{f_{f_8}}{\sqrt{6}} \frac{m_{f_8}}{ m_{f_{1}(1285)}}\sin\theta_{^{3}P_{1}},  \nonumber \\
&& f^{u}_{f_{1}(1420)} = - \frac{f_{f_1}}{\sqrt{3}} \frac{m_{f_1}}{ m_{f_{1}(1420)}}\sin\theta_{^{3}P_{1}} + \frac{f_{f_8}}{\sqrt{6}} \frac{m_{f_8}}{ m_{f_{1}(1420)}}\cos
\theta_{^{3}P_{1}},
\label{dc2}
\end{eqnarray}

\noindent and
\begin{eqnarray}
&& f^{s}_{f_{1}(1285)} = \frac{f_{f_1}}{\sqrt{3}} \frac{m_{f_1}}{ m_{f_{1}(1285)}}\cos\theta_{^{3}P_{1}} - \frac{2\;f_{f_8}}{\sqrt{6}} \frac{m_{f_8}}{ m_{f_{1}(1285)}}\sin\theta_{^{3}P_{1}},  \nonumber \\
&& f^{s}_{f_{1}(1420)} = - \frac{f_{f_1}}{\sqrt{3}} \frac{m_{f_1}}{ m_{f_{1}(1420)}}\sin\theta_{^{3}P_{1}} - \frac{2\;f_{f_8}}{\sqrt{6}} \frac{m_{f_8}}{ m_{f_{1}(1420)}}\cos
\theta_{^{3}P_{1}}.
\label{dc2}
\end{eqnarray}

The mixing angle $\theta_{^{3}P_{1}}$ has been calculated theoretically in some references (see for example \cite{axialQCDF1,mixing1}). The Ref. \cite{mixing1} found that this mixing angle has two values: $\theta_{^{3}P_{1}} = 38^{\circ},   50^{\circ}$. The previous phenomenological analysis did in Ref. \cite{Close} suggests that $\theta_{^{3}P_{1}} \simeq 50^{\circ}$.\\

In this work we use the predictions of Ref. \cite{axialQCDF2} for decay constants of  $f_{1}(1285)$ and $f_{1}(1420)$ mesons (see Table III). These values were calculated using the Gell-Mann-Okubo mass formula and the value $\theta_{K_{1}} = -37^{\circ}$ $(-58^{\circ})$ for the $K_{1A}-K_{1B}$ mixing angle. It was found that the mixing angle for  ${}^{3}P_{1}$ states is $\theta_{^{3}P_{1}} = 27.9^{\circ}$ $(53.2^{\circ})$. So, we can see that the mixing angle $\theta_{^{3}P_{1}}$ depends on the angle $\theta_{K_{1}}$. If the mixing were ideal, the $f_{1}(1285)$ meson  will be  made
up of $(u\bar{u}+d\bar{d})/2$ while $f_{1}(1420)$ is composed of $s\bar{s}$.\\

\begin{table}[ht]
{\small Table III. Decay constants $f^{q}_{f_{1}(1285)}$ and $f^{q}_{f_{1}(1420)}$}
\par
\begin{center}
\renewcommand{\arraystretch}{1.2}
\renewcommand{\arrayrulewidth}{0.8pt}
\begin{tabular}{lcc}
\hline\hline
 $f_{^{3}P_{1}}$(MeV) & $\theta_{^{3}P_{1}} = 53.2^{\circ}$ & $\theta_{^{3}P_{1}} = 27.9^{\circ}$\\
 \hline
 $f^{u}_{f_{1}(1285)}$ & 172 & 178\\
 $f^{u}_{f_{1}(1420)}$ & $-55$ & 23 \\
 $f^{s}_{f_{1}(1285)}$ & $-72$ & 29\\
 $f^{s}_{f_{1}(1420)}$ & $-219$ & $-230$\\
\hline\hline
\end{tabular}%
\end{center}
\end{table}

On the other hand, for the $a_{1}(1260)$ decay constant we have taken $f_{a_1}$= 238 MeV obtained
using the QCD sum rule method \cite{mixing1}.\\

We use the parametrization of the CKM matrix  in terms of Wolfenstein parameters $\lambda
$, $A$, $\bar\rho$ and $\bar\eta$ \cite{wolfenstein}:

\begin{equation}
V_{\text{CKM}} = \left(%
\begin{array}{ccc}
1-{\frac{1}{2}}\lambda^2 & \lambda & A\lambda^3(\rho-i\eta) \\
-\lambda & 1-{\frac{1}{2}}\lambda^2 & A\lambda^2 \\
A\lambda^3(1-\rho-i\eta) & -A\lambda^2 & 1%
\end{array}%
\right) + \mathcal{O}(\lambda^4),
\end{equation}

\noindent with $\rho = \bar\rho(1-\lambda^2/2)^{-1}$ and $\eta=\bar\eta(1- \lambda^2/2)^{-1}$. We take  central values from the global fit for Wolfenstein parameters: $\lambda =$ 0.2257, $A$ = 0.814, $\bar\rho$ = 0.135 and $\bar\eta$ = 0.349 \cite{PDG}. \\

Masses and average lifetimes of neutral and charged $B$ mesons were taken from \cite{PDG}. The running quark masses are given at the scale $\mu \approx m_b$, since the energy released in $B$
decays is of the order of $m_b$. We use $m_u(m_b)=3.2$ MeV, $m_d(m_b)=6.4$ MeV, $%
m_s(m_b)=127$ MeV, $m_c(m_b)=0.95$ GeV and $m_b(m_b)=4.34$ GeV (see Ref. \cite{fusaoku}).\\

\section{Numerical results and discussion}

In this section, we present our   numerical values for  branching ratios of nonleptonic charmless $B \rightarrow A(^3P_1)T$ decays, using $B \to T$ form factors obtained in the CLF approach \cite{Cheng04}. Also, we establish a comparison between $B \to AT$ and $B \to VT$ modes.\\

The decay width for the $B \to  A T$ process is given by
\begin{align}\label{AT}
\Gamma(B \to  A T) =& \frac{G_{F}^{2}}{32\pi m_{B}^{3}} f_{A}^{2} \bigg|V_{ub}^{*}V_{ud(s)} \cdot a_{1,2}  - V_{tb}^{*}V_{td(s)} \cdot (\Phi_{penguin})\bigg|^{2} \\ \nonumber
& \times \left[ \alpha(m^{2}_{A})\lambda^{7/2} + \beta(m^{2}_{A})\lambda^{5/2}+ \gamma(m^{2}_{A})\lambda^{3/2} \right],
\end{align}

\noindent where we have summed over polarizations of the tensor meson $T$. The $\Phi_{penguin}$ factor is a linear combination of penguin coefficients $a_{3}, ..., a_{10}$, $\lambda \equiv \lambda(m^{2}_{B},m^{2}_{T},m^{2}_{A})= (m^{2}_{B}+m^{2}_{T}-m^{2}_{A})^{2} - 4 m^{2}_{B}m^{2}_{T}$ is the triangle function, and  $\alpha$, $\beta$
and $\gamma$ are quadratic functions of  form factors $k$,
$b_{+}$ and $h$, evaluated at $q^2 = m^{2}_{A}$. They are expressed by

\begin{equation}
\alpha(m^{2}_{A}) = \frac{b_{+}^{2}}{24m_{T}^{4}},
\end{equation}

\begin{equation}
\beta(m^{2}_{A}) = \frac{1}{24m_{T}^{4}} \left[k^{2}+6m_{T}^{2}m^{2}_{A}h^{2}+2(m_{B}^{2}-m_{T}^{2}-m^{2}_{A})kb_{+}\right],
\end{equation}

\begin{equation}
\gamma(m^{2}_{A}) = \frac{5 m^{2}_{A}k^{2}}{12m_{T}^{2}}.\\
\end{equation}

The ratio between  decay widths of  $B \to AT$ (see Eq. (\ref{AT})) and $B \to VT$ (see Ref. \cite{Munoz}) channels, where $A$ and $V$ mesons have the same quark content, is

\begin{equation}\label{ratio}
\mathcal{R}_{AT/VT}  = \frac{\Gamma(B \to AT)}{\Gamma(B \to VT)} = \bigg( \frac{f_A}{f_V} \bigg)^{2} \Bigg[ \frac{\mathcal{Z}_{A}^{QCD}}{\mathcal{Z}_{V}^{QCD}} \Bigg]^{2} \left[\frac{ \alpha(m^{2}_{A})\lambda_{A}^{7/2} + \beta(m^{2}_{A})\lambda_{A}^{5/2}+ \gamma(m^{2}_{A})\lambda_{A}^{3/2}}{ \alpha(m^{2}_{V})\lambda_{V}^{7/2} + \beta(m^{2}_{V})\lambda_{V}^{5/2}+ \gamma(m^{2}_{V})\lambda_{V}^{3/2}} \right]
\end{equation}

\noindent where $\lambda_{A(V)}  = \lambda(m^{2}_{B},m^{2}_{T},m^{2}_{A(V)})$ and $\mathcal{Z}_{A(V)}^{QCD}$ is a sum of products of CKM elements with QCD coefficients $a_i$ ($i=1,..., 10$). We can see that $\mathcal{R}_{AT/VT}$ is conformed by the product of three terms: the first one is the ratio between decay constants $f_A$ and $f_V$; the second factor  corresponds to the ratio between QCD contributions; and the third term comes from form factors and  kinematical $\lambda_{A(V)}$ function. This ratio can be considered as a test of the factorization approximation. If  $a_i$ coefficients and form factors are known, decay constants  can be determined  from $\mathcal{R}_{AT/VT}$. On the other hand, $\mathcal{R}_{AT/VT}$ is a test of form factors if decay constants and $a_i$ coefficients are known.\\

The QCD contributions for $B^{+,0} \to a_{1}^{+}a_{2}^{0,-}$ and $B^{+,0} \to \rho^{+} a_{2}^{0,-}$ modes are the same,  i.e., the ratio  $\mathcal{Z}_{A}^{QCD}/ \mathcal{Z}_{V}^{QCD} = 1$.  A similar situation occurs with decays  $B^{+,0} \to \bar{K_{1}}^{0}K_{2}^{*+,0}$ and $B^{+,0} \to \bar{K_{}}^{*0}K_{2}^{*+,0}$, and $B \to K_{1}a_{2}$ and $B \to K^{*}a_{2}$.  In these cases,  $\mathcal{R}_{AT/VT}$ gives a better test of factorization scheme. If decay constants are known $\mathcal{R}_{AT/VT}$ can give a test of form factors.\\

For obtaining  branching ratios of exclusive charmless $B \rightarrow A(^3P_1)T$ decays, we have taken  expressions for  decay amplitudes  given in  Appendix. These expressions include all  contributions of $H_{eff}$.  Our numerical results   are listed in Tables IV and V. Specifically, branching fractions of $B \to K_{1}T$ (whit $K_{1}=K_{1}(1270), K_{1}(1400)$) modes are shown in Table IV, and  predictions for branchings of  $B \to A(^3P_1)T$, where $A$ is a nonstrange axial-vector meson, are collected in Table V.\\

\begin{table}[ht]
{\small Table IV.~Branching ratios (in units of $10^{-6}$) for charmless $B \to K_{1}T$ decays, where $K_{1}=K_{1}(1270)$,  $K_{1}(1400)$ is a strange axial-vector meson, evaluated in two different numerical values of the mixing angle $\theta_{K_{1}}$.}
\par
\begin{center}
\renewcommand{\arraystretch}{1.2}
\renewcommand{\arrayrulewidth}{0.8pt}
\begin{tabular}{lccc}
\hline\hline
    & & $\mathcal{B}$ & \\
    \cline{2-4}
   Process  & $ -37^{\circ}$ &&  $-58^{\circ}$  \\
  \hline
  $B^{+} \to K_{1}(1270)^{+}a_{2}^{0}$ & 1.88 && 3.04  \\
  $B^{+} \to K_{1}(1400)^{+}a_{2}^{0}$ &  1.70 && 5.43 \\
  $B^{0} \to K_{1}(1270)^{+}a_{2}^{-}$ &  3.51 && 5.68 \\
  $B^{0} \to K_{1}(1400)^{+}a_{2}^{-}$ &  3.18 && 1.02 \\
  &  \\
  Penguin process & & & \\
  \hline
  $B^{+} \to \bar{K_{1}}(1270)^{0}K_{2}^{*+}$ & 0.37 && 0.60 \\
  $B^{+} \to \bar{K_{1}}(1400)^{0}K_{2}^{*+}$ & 0.33 && 0.11 \\
  $B^{+} \to K_{1}(1270)^{0}a_{2}^{+}$ & 5.77 && 9.34 \\
  $B^{+} \to K_{1}(1400)^{0}a_{2}^{+}$ & 5.24 && 1.67 \\
  $B^{0} \to \bar{K_{1}}(1270)^{0}K_{2}^{*0}$ & 0.34 && 0.55 \\
  $B^{0} \to \bar{K_{1}}(1400)^{0}K_{2}^{*0}$ & 0.30 && 0.09 \\
  $B^{0} \to K_{1}(1270)^{0}a_{2}^{0}$ & 2.70 && 4.38 \\
  $B^{0} \to K_{1}(1400)^{0}a_{2}^{0}$ & 2.48 && 0.78 \\
  \hline\hline
\end{tabular}
\end{center}
\end{table}

\begin{table}[ht]
{\small Table V.~Branching ratios (in units of $10^{-6}$) for charmless $B \to A(^3P_1)T$ decays, where $A$ is a nonstrange axial-vector meson. Processes including $f_{1}(1285)$ and $f_{1}(1420)$ mesons are considered with two different values of mixing angle $\theta_{^{3}P_{1}} = 53.2^{\circ}\; [27.9^{\circ}]$.}
\par
\begin{center}
\renewcommand{\arraystretch}{1.2}
\renewcommand{\arrayrulewidth}{0.8pt}
\begin{tabular}{lc}
\hline\hline
   Process &  $\mathcal{B}$  \\
  \hline
  $B^{+} \to a_{1}^{+}a_{2}^{0}$ & 22.71 \\
  $B^{+} \to a_{1}^{0}a_{2}^{+}$ & 0.085 \\
  $B^{+} \to f_{1}(1285)a_{2}^{+}$ & 0.17\;[0.12] \\
  $B^{+} \to f_{1}(1420)a_{2}^{+}$ & 0.02\;[0.06] \\
  $B^{+} \to a_{1}^{0}K_{2}^{*+}$ & 0.77 \\
  $B^{+} \to f_{1}(1285)K_{2}^{*+}$ & 3.12\;[0.61] \\
  $B^{+} \to f_{1}(1420)K_{2}^{*+}$ & 4.02\;[6.42] \\
  $B^{0} \to a_{1}^{+} a_{2}^{-}$ & 42.47 \\
  $B^{0} \to a_{1}^{0}a_{2}^{0}$ & 0.04 \\
  $B^{0} \to f_{1}(1285)a_{2}^{0}$ & 0.08\;[0.06] \\
  $B^{0} \to f_{1}(1420)a_{2}^{0}$ & 0.009\;[0.03] \\
  $B^{0} \to a_{1}^{0}K_{2}^{*0}$ & 0.71 \\
  $B^{0} \to f_{1}(1285)K_{2}^{*0}$ & 2.87\;[0.56] \\
  $B^{0} \to f_{1}(1420)K_{2}^{*0}$ & 3.70\;[5.91] \\

  \hline\hline
\end{tabular}%
\end{center}
\end{table}

Branching ratios of $B^{0} \to a_{1}^{+}a_{2}^{-}$ and $B^{+} \to a_{1}^{+}a_{2}^{0}$  modes are the biggest. They are $42.47 \times10^{-6}$ and $22.71 \times10^{-6}$, respectively (see Table V). These decays receive contributions of the $a_1$ coefficient and the linear combination $a_4 +a_{10}$ (see Appendix).  $B^{+} \to K_{1}^{+}a_{2}^{0}$ and $B^{0} \to K_{1}^{+}a_{2}^{-}$ modes, with $K_{1}=K_{1}(1270), K_{1}(1400)$, also have  sizable branching ratios of $(1.7 - 3.5) \times 10^{-6}$ and $(1 - 5.6) \times 10^{-6}$ with $\theta_{K_{1}} = -37^{\circ}$ and  $\theta_{K_{1}} = -58^{\circ}$, and receive contribution of  same QCD coefficients with different CKM matrix elements, respectively (see Table IV).  Another feature is that branching fractions of $B \to K_{1}(1270)a_{2}$ and $B \to K_{1}(1400)a_{2}$ are almost equal for $\theta_{K_{1}} = -37^{\circ}$ while are different for $\theta_{K_{1}} = -58^{\circ}$. Thus, the measurement of the ratio $\mathcal{B}(B \to K_{1}(1270)a_{2})/\mathcal{B}(B \to K_{1}(1400)a_{2})$ will be a test of the value of the mixing angle $\theta_{K_{1}}$.\\

 For  color-suppressed decays, $B \to f_{1}K_{2}^{*}$ modes, with $f_{1}=f_{1}(1285), f_{1}(1420)$, have the biggest branching ratios (see Table V).  They are  $(2.8 - 4)\times10^{-6}$ with  the mixing angle $\theta_{^{3}P_{1}} = 53.2^{\circ}$. On the other hand, if $\theta_{^{3}P_{1}} = 27.9^{\circ}$, only $B \to f_{1}(1420)K_{2}^{*}$ decays have branching ratios of $10^{-6}$.  We can see that $\mathcal{B}(B \to f_1(1420)K_{2}^{*}) \approx (1.3)\mathcal{B}(B \to f_1(1285)K_{2}^{*})$ and  $\mathcal{B}(B \to f_1(1420)K_{2}^{*}) \approx (10.5)\mathcal{B}(B \to f_1(1285)K_{2}^{*})$ with $\theta_{^{3}P_{1}} = 53.2^{\circ}$ and $\theta_{^{3}P_{1}} = 27.9^{\circ}$, respectively. Thus, the measurement of the ratio  $\mathcal{B}(B \to f_1(1420)K_{2}^{*})/\mathcal{B}(B \to f_1(1285)K_{2}^{*})$ will help to determine the mixing angle $\theta_{^{3}P_{1}}$.\\

 $B \to K_1^{0}K_2^{*}(a_2)$ decays (see Table IV) are  penguin-dominated and   receive contribution of the linear combination $(a_4 - a_{10}/2)$.  $B \to K_{1}^{0}a_{2}$ processes, with $K_1 = K_1(1270), K_1(1400)$, have branching ratios of $(2.4 - 5.7) \times 10^{-6}$ with $\theta_{K_{1}} =  -37^{\circ}$, while their branching fractions are  $(0.7 - 9.34) \times10^{-6}$  if $\theta_{K_{1}} =  -58^{\circ}$. Another interesting relation is that
 \begin{equation}
 \mathcal{B}(B \to K_1^{0}(1270)K_2^{*}(a_2))|_{\theta_{K_{1}} =  -37^{\circ}} < \mathcal{B}(B \to K_1^{0}(1270)K_2^{*}(a_2))|_{\theta_{K_{1}} =  -58^{\circ}}
 \end{equation}
 while
 \begin{equation}
 \mathcal{B}(B \to K_1^{0}(1400)K_2^{*}(a_2))|_{\theta_{K_{1}} =  -37^{\circ}} > \mathcal{B}(B \to K_1^{0}(1400)K_2^{*}(a_2))|_{\theta_{K_{1}} =  -58^{\circ}}.
 \end{equation}
 Moreover, branching ratios of $B \to K_1^{0}a_2$ decays are insensitive to the mixing angle ${\theta_{K_{1}} =  -37^{\circ}}$. This behavior  is opposite for  ${\theta_{K_{1}} =  -58^{\circ}}$. Hence, the ratio $\mathcal{B}(B \to K_{1}(1270)^0a_{2})/\mathcal{B}(B \to K_{1}(1400)^0a_{2})$  can offer a better determination for $\theta_{K_{1}}$. Decay rates of $B \to K_1^{0}K_2^{*}$ channels are small because they arise from $b \to d$ penguin transition and are suppressed by  the smallness of  respective Wilson coefficients.  \\

We have compared branching ratios of  $B \to AT$ decays (obtained in this work) and $B \to VT$ channels (obtained recently for us using the CLF approach\cite{tensor1}), where $A$ and $V$ mesons have the same quark content. In general, the ratio $\mathcal{R}_{AT/VT} \gtrsim 1$ when  $A$, $V$ and $T$ mesons are non-strange. It means that the product of the second and the third factors in Eq. (\ref{ratio}) is approximately 1 and that  $\mathcal{B}(B \to a_1a_2)/\mathcal{B}(B \to \rho a_2) \approx (f_{a_1}/f_{\rho})^2$. This ratio can be at  the reach of $B$ factories and LHC-b experiment.

%*********************************************************************************************
%*********************************************************************************************
\newpage
\section{Conclusions}

In this work we study the production of excited orbitally ($p$-wave) mesons in  nonleptonic charmless two-body $B$ decays. We compute branching ratios of  $B\to A(^{3}P_{1})T$ decays within the framework of generalized factorization, using  form factors from CLF approach for $B \to T$ transitions. Respective factorized amplitudes of these decays are explicitly showed in Appendix. We obtained that $\mathcal{B}(B^{0} \to a_{1}^{+}a_{2}^{-}) =42.47 \times10^{-6}$, $\mathcal{B}(B^{+} \to a_{1}^{+}a_{2}^{0}) = 22.71 \times10^{-6}$, $\mathcal{B}(B \to f_{1}K_{2}^{*}) = (2.8-4) \times 10^{-6}$  (with $f_{1}=, f_{1}(1285),f_{1}(1420)$) for $\theta_{^{3}P_{1}} = 53.2^{\circ}$, $\mathcal{B}(B \to f_{1}(1420)K_{2}^{*}) = (5.91-6.42) \times 10^{-6}$ with $\theta_{^{3}P_{1}} = 27.9^{\circ}$, $\mathcal{B}(B \to K_{1}a_{2})= (1.7 - 5.7) [1-9.3]  \times10^{-6}$ for $\theta_{K_{1}} =  -37^{\circ} [-58^{\circ}]$ where $K_1 = K_1(1270), K_1(1400)$. It seems that the majority of these modes could be measured at the present asymmetric $B$ factories, BABAR and Belle, as well as at future hadronic $B$ experiments such as BTeV and LHC-b. Additionally, we have found that   $\mathcal{B}(B \to K_{1}(1270)a_{2})/\mathcal{B}(B \to K_{1}(1400)a_{2})$ and $\mathcal{B}(B \to f_1(1420)K_{2}^{*})/\mathcal{B}(B \to f_1(1285)K_{2}^{*})$ ratios will help to determine  mixing angles $\theta_{K_{1}}$ and $\theta_{^{3}P_{1}}$, respectively.

%*********************************************************************************************
%*********************************************************************************************

\begin{acknowledgments}
The authors acknowledge financial support to
{\it Comit\'e Central de Investigaciones} of University of Tolima.
\end{acknowledgments}

%*********************************************************************************************
%*********************************************************************************************

\appendix*

\begin{center}
\section{Decay Amplitudes}
\end{center}
In this appendix, we present expressions for the factorizable decay amplitudes of charmless $B \to A(^{3}P_{1})T$ decays. They must be multiplied by $G_F\epsilon^{*\mu\nu}/\sqrt{2}$.\\

\subsection{Process $|\Delta S|=0$}

\begin{align}\label{}
\mathcal{A}(B^{+} \to a_{1}^{+}a_{2}^{0}  )=& \frac{1}{\sqrt{2}} m_{a_{1}}f_{a_{1}} F_{\mu\nu}^{B \to a_{2}}(m^{2}_{a_{1}}) \Bigg\{V_{ub}^{*}V_{ud} a_{1} - V_{tb}^{*}V_{td}(a_{4}+a_{10})\Bigg\}
\end{align}

\begin{align}\label{}
\mathcal{A}(B^{+} \to a_{1}^{0} a_{2}^{+}  )=& \frac{1}{\sqrt{2}} m_{a_{1}}f_{a_{1}} F_{\mu\nu}^{B \to a_{2}}(m^{2}_{a_{1}}) \Bigg\{V_{ub}^{*}V_{ud} a_{2} - V_{tb}^{*}V_{td}\Bigg[-a_{4} + \frac{1}{2}(-3a_{7}+3a_{9}+a_{10}) \Bigg]\Bigg\}
\end{align}

\begin{align}\label{}
\mathcal{A}(B^{+} \to f_{1} a_{2}^{+} )= m_{f_{1}}f^{u}_{f_{1}} F_{\mu\nu}^{B \to a_{2}}(m^{2}_{f_{1}})
\Bigg\{ & V_{ub}^{*}V_{ud} a_{2} - V_{tb}^{*}V_{td} \Bigg[2(a_{3}-a_{5})+a_{4} - \frac{1}{2}(a_{7}-a_{9}+a_{10}) \nonumber\\
&+ \left((a_{3}-a_{5}) + \frac{1}{2}(a_{7}-a_{9}) \right)  \Bigg(\frac{f^{s}_{f_{1}}}{f^{u}_{f_{1}}}\Bigg) \Bigg]\Bigg\}
\end{align}

\begin{align}\label{}
\mathcal{A}(B^{+} \to \bar{K_{1}^{0}} K_{2}^{*+} )=& m_{K_{1}}f_{K_{1}}  F_{\mu\nu}^{B \to K_{2}^{*}}(m^{2}_{K_{1}})  \Bigg\{- V_{tb}^{*}V_{td} \Bigg[a_{4} - \frac{1}{2}a_{10} \Bigg]\Bigg\}
\end{align}

\begin{align}\label{}
\mathcal{A}(B^{0} \to a_{1}^{+}a_{2}^{-}  )=& m_{a_{1}}f_{a_{1}} F_{\mu\nu}^{B \to a_{2}}(m^{2}_{a_{1}}) \Bigg\{V_{ub}^{*}V_{ud} a_{1} - V_{tb}^{*}V_{td}(a_{4}+a_{10})\Bigg\}
\end{align}

\begin{align}\label{}
\mathcal{A}(B^{0} \to a_{1}^{-} a_{2}^{+}  )= 0.
\end{align}

\begin{align}\label{}
\mathcal{A}(B^{0} \to a_{1}^{0}a_{2}^{0}  )=& \frac{1}{2} m_{a_{1}}f_{a_{1}}  F_{\mu\nu}^{B \to a_{2}}(m^{2}_{a_{1}}) \Bigg\{V_{ub}^{*}V_{ud} a_{2} - V_{tb}^{*}V_{td}\Bigg[-a_{4} + \frac{1}{2}(-3a_{7}+3a_{9}+a_{10}) \Bigg]\Bigg\}
\end{align}

\begin{align}\label{}
\mathcal{A}(B^{0} \to f_{1}a_{2}^{0} )= \frac{1}{\sqrt{2}}m_{f_{1}}f^{u}_{f_{1}} F_{\mu\nu}^{B \to a_{2}}(m^{2}_{f_{1}})
\Bigg\{ & V_{ub}^{*}V_{ud} a_{2} - V_{tb}^{*}V_{td}\Bigg[2(a_{3}-a_{5})+a_{4} - \frac{1}{2}(a_{7}-a_{9}+a_{10}) \nonumber\\
&+ \Bigg((a_{3}-a_{5}) + \frac{1}{2}(a_{7}-a_{9})\Bigg)\Bigg(\frac{f^{s}_{f_{1}}}{f^{u}_{f_{1}}}\Bigg) \Bigg]\Bigg\}
\end{align}

\begin{align}\label{}
\mathcal{A}(B^{0} \to \bar{K_{1}^{0}} K_{2}^{*0} )=& m_{K_{1}}f_{K_{1}} F_{\mu\nu}^{B \to K_{2}^{*}}(m^{2}_{K_{1}})  \Bigg\{- V_{tb}^{*}V_{td}\Bigg[a_{4} - \frac{1}{2}a_{10} \Bigg]\Bigg\}
\end{align}

\subsection{Process $|\Delta S|=1$}

\begin{align}\label{}
\mathcal{A}(B^{+} \to K_{1}^{+}a_{2}^{0})=& \frac{1}{\sqrt{2}} m_{K_{1}}f_{K_{1}}  F_{\mu\nu}^{B \to a_{2}}(m^{2}_{K_{1}}) \Bigg\{V_{ub}^{*}V_{us} a_{1} - V_{tb}^{*}V_{ts}(a_{4}+a_{10})\Bigg\}
\end{align}

\begin{align}\label{}
\mathcal{A}(B^{+} \to a_{1}^{0}K_{2}^{*+}  )=& \frac{1}{\sqrt{2}} m_{a_{1}}f_{a_{1}} F_{\mu\nu}^{B \to K_{2}^{*}}(m^{2}_{a_{1}}) \Bigg\{V_{ub}^{*}V_{us} a_{2} - V_{tb}^{*}V_{ts}\Bigg[\frac{3}{2}(-a_{7}+a_{9}) \Bigg]\Bigg\}
\end{align}

\begin{align}\label{}
\mathcal{A}(B^{+} \to f_{1}K_{2}^{*+})=  m_{f_{1}}f^{u}_{f_{1}} F_{\mu\nu}^{B \to K_{2}^{*}}(m^{2}_{f_{1}}) \Bigg\{ & V_{ub}^{*}V_{us} a_{2} - V_{tb}^{*}V_{ts}\Bigg[2(a_{3}-a_{5})- \frac{1}{2}(a_{7}-a_{9}) \nonumber\\
&+ \Bigg((a_{3}+a_{4}-a_{5}) + \frac{1}{2}(a_{7}-a_{9}-a_{10})\Bigg)\Bigg(\frac{f^{s}_{f_{1}}}{f^{u}_{f_{1}}}\Bigg) \Bigg]\Bigg\}
\end{align}

\begin{align}\label{}
\mathcal{A}(B^{+} \to K_{1}^{0}a_{2}^{+} )=& m_{K_{1}}f_{K_{1}}  F_{\mu\nu}^{B \to a_{2}}(m^{2}_{K_{1}})  \Bigg\{- V_{tb}^{*}V_{ts} \Bigg[a_{4} - \frac{1}{2}a_{10} \Bigg]\Bigg\}
\end{align}

\begin{align}\label{}
\mathcal{A}(B^{0} \to K_{1}^{+}a_{2}^{-} )=&  m_{K_{1}}f_{K_{1}} \epsilon^{*\mu\nu} F_{\mu\nu}^{B \to a_{2}}(m^{2}_{K_{1}}) \Bigg\{V_{ub}^{*}V_{us} a_{1} - V_{tb}^{*}V_{ts}(a_{4}+a_{10})\Bigg\}
\end{align}

\begin{align}\label{}
\mathcal{A}(B^{0} \to a_{1}^{0}K_{2}^{*0} )=& \frac{1}{\sqrt{2}} m_{a_{1}}f_{a_{1}}  F_{\mu\nu}^{B \to K_{2}^{*}}(m^{2}_{a_{1}}) \Bigg\{V_{ub}^{*}V_{us}a_{2} - V_{tb}^{*}V_{ts}\Bigg[\frac{3}{2}(-a_{7}+a_{9}) \Bigg]\Bigg\}
\end{align}

\begin{align}\label{}
\mathcal{A}(B^{0} \to f_{1}K_{2}^{*0})= m_{f_{1}}f^{u}_{f_{1}} F_{\mu\nu}^{B \to K_{2}^{*}}(m^{2}_{f_{1}}) \Bigg\{ & V_{ub}^{*}V_{us} a_{2} - V_{tb}^{*}V_{ts} \Bigg[2(a_{3}-a_{5})- \frac{1}{2}(a_{7}-a_{9}) \nonumber\\
&+ \Bigg((a_{3}+a_{4}-a_{5}) + \frac{1}{2}(a_{7}-a_{9}-a_{10})\Bigg)\Bigg(\frac{f^{s}_{f_{1}}}{f^{u}_{f_{1}}}\Bigg) \Bigg]\Bigg\}
\end{align}

\begin{align}\label{}
\mathcal{A}(B^{0} \to K_{1}^{0}a_{2}^{0})=& \frac{1}{\sqrt{2}} m_{K_{1}}f_{K_{1}}  F_{\mu\nu}^{B \to a_{2}}(m^{2}_{K_{1}})  \Bigg\{- V_{tb}^{*}V_{ts} \Bigg[a_{4} - \frac{1}{2}a_{10} \Bigg]\Bigg\}
\end{align}

\noindent with
\begin{equation}\label{}
F_{\mu\nu}^{B \to T}(m_{A}^{2}) \equiv  \epsilon^{*}_{\alpha} (p_{B}+p_{T})_{\rho} \left[i h(m_{A}^{2}) \cdot \varepsilon^{\alpha\beta\rho\sigma} g_{\mu\beta}(p_{A})_{\nu} (p_{A})_{\sigma}+ k(m_{A}^{2})\cdot \delta^{\alpha}{}_{\mu}\delta^{\rho}{}_{\nu}+b_{+}(m_{A}^{2}) \cdot(p_{A})_{\mu} (p_{A})_{\nu} g^{\alpha\rho} \right],
\end{equation}

\noindent where $T$ stands for $a_{2}$ and $K_{2}^{*}$.

%*********************************************************************************************
%*********************************************************************************************

%*********************************************************************************************
%\newpage

\end{document}